\documentclass[conference]{IEEEtran}

\usepackage{color}
\usepackage{amsmath}
\usepackage{mathtools}
\usepackage[]{algorithm2e}
\usepackage{algorithmic}

\algsetup{linenosize=\small}
\usepackage[table,xcdraw]{xcolor}
\usepackage{multirow}
\usepackage[super]{nth}
\usepackage{graphicx}
\usepackage{caption}
\usepackage{subcaption}
\usepackage{textcomp}
\usepackage{xspace}
\usepackage{siunitx}
\usepackage{epsfig}
\usepackage{epstopdf}
\usepackage{soul}
\usepackage{url}
\usepackage{tablefootnote}
\usepackage[labelformat=simple]{subcaption}
\let\OldTexttrademark\texttrademark
\renewcommand{\texttrademark}{\OldTexttrademark\xspace}%
\begin{document}
	
	\title{Learning Optimal Routing for the Uplink in LPWANs Using Similarity-enhanced $\epsilon$-greedy}
	
	% COMMENTS title: Optimal LPWAN learning multi-hop uplink Reinforcement Learning (fora energy-aware) Similarity
	
	\author{\IEEEauthorblockN{Sergio Barrachina-Mu\~noz and Boris Bellalta}
		\IEEEauthorblockA{Dept. of Information and Communication Technologies\\
			Universitat Pompeu Fabra, Barcelona (Spain)\\
			Email: sergio.barrachina@upf.edu, boris.bellalta@upf.edu}}
	
	\maketitle
	
	%--------------------
	%--------------------
	% ABSTRACT
	%--------------------
	%--------------------
	
	\begin{abstract}
		
		% IoT and how LPWANs may contribute
		Despite being a relatively new communication technology, Low-Power Wide Area Networks (LPWANs) have shown their suitability to empower a major part of Internet of Things applications.
		%due to their energy-aware designs, large coverage areas, low cost, and scalability.
		Nonetheless, most LPWAN solutions are built on star topology (or single-hop) networks, often causing lifetime shortening in stations located far from the gateway. In this respect, recent studies show that multi-hop routing for uplink communications can reduce LPWANs' energy consumption significantly. However, it is a troublesome task to identify such energetically optimal routing through trial-and-error brute-force approaches because of time and, especially, energy consumption constraints.
		
		% What we do in this work
 		 In this work we show the benefits of facing this exploration/exploitation problem by running centralized variations of the multi-arm bandit's $\epsilon$-greedy, a well-known online decision-making method that combines best known action selection and knowledge expansion. Important energy savings are achieved when proper randomness parameters are set, which are often improved when conveniently applying \textit{similarity}, a concept introduced in this work that allows harnessing the gathered knowledge by sporadically selecting unexplored routing combinations akin to the best known one.
		
	\end{abstract}

	%--------------------
	%--------------------
	% INTRODUCTION
	%--------------------
	%--------------------
	
	\section{Introduction} \label{sec:introduction}
	
	% LPWAN arise as a solution to IoT
	Low-Power Wide Area Networks (LPWANs) are wireless wide area networks designed for achieving extensive coverage ranges, extending end devices battery lifetime and reducing the operational cost of traditional cellular networks. Consequently, they are considered to be a promising complementary communication technology for a broad range of Internet of Things (IoT) applications. LPWANs are characterized by exploiting the sub-1GHz unlicensed, industrial, scientific and medical (ISM) frequency band, and by sporadically transmitting small packets at low data rates, which enables remarkably low receptor sensitivities.
	
	% LPWANs and single-hop
	Nonetheless, most LPWAN solutions like LoRaWAN\texttrademark \cite{loraWanSpec} or SIGFOX\texttrademark \cite{sigfox2016main} are built following star topologies, where end devices, or stations (STAs), are connected directly to the base station or gateway (GW), making those ones located far from the GW to rely deeply on their transceiver's capabilities. This may lead to high energy consumption in such STAs, while hindering the inclusion of devices with transmission power limitations because of the aforesaid range constraint. Therefore, even though star topologies have clear benefits like the robustness achieved by centralized management, they may not be energy-efficient depending on the scenario. In that sense, despite there is scarce literature on studies comparing the energy consumption of single-hop and multi-hop in this kind of networks, authors in \cite{adame2017hare} state that HARE, a novel LPWAN protocol stack enabling multi-hop communication in the uplink, has proven energy savings of up to 15\% in a real testbed. Likewise, authors in \cite{barrachina2016multi} use the \textit{Distance-Ring Exponential Station Generator} (DRESG)\footnote{All of the source code in DRESG framework for LPWANs is open \cite{barrachina2017dresg}, encouraging sharing of algorithms between contributors and providing the ability for people to improve on the work of others under the GNU General Public License v3.0.} analytical framework for showing that, in LPWANs of up to several thousands of STAs, multi-hop leads to higher network lifetimes than single-hop since the consumption of STAs located far from the GW is significantly reduced. 
	
	% Lack of studies of multi-hop LPWANs, DRESG and HARE
	%Although there are few LPWAN solutions enabling multi-hop (or tree) topologies like DASH7 \cite{weyn2015dash7}, there is scarce literature on studies comparing the energy consumption impact of single-hop and multi-hop approaches in this kind of networks.
	
	% Lack of studies estimating the energy required to identify energy-aware routing in LPWANs.
	Still, authors in \cite{barrachina2016multi} do not provide any estimation on the energy consumption required to identify the optimal routing in a centralized way, but try all the possible routing combinations in a brute-force approach, which may be critical depending on the amount of such combinations. Therefore, as an exploration/exploitation problem, reinforcement learning (RL) algorithms like multi-armed bandits' $\epsilon$-greedy are appropriate due to their online decision-making, which combines selecting the best known routing (i.e, exploiting), and extending the gathered knowledge (i.e, exploring).
	
	% Paper contributions and Remainder of paper organization
	In this work we study the energy savings achieved when learning the optimal routing through the well-known $\epsilon$-greedy method, and through a \textit{similarity-enhanced} extended version. The main contributions are described below:
	
	\begin{itemize}
		\item Characterization of the effect on energy consumption of $\epsilon$-greedy's settings (initial value and updating function).
		\item Definition of the \textit{similarity} concept for harnessing the gathered knowledge in online decision-making.
		\item Energy consumption characterization and appropriate use-cases identification of \textit{similarity-enhanced} $\epsilon$-greedy.
	\end{itemize}

	Results show that the best settings of $\epsilon$-greedy are really tied to the LPWAN deployment, and that exploiting gathered knowledge can significantly improve their performance.

	\section{DRESG framework for LPWANs} \label{sec:dresg}
	
	 DRESG is a framework that allows estimating the energy consumption of every node in a network given a certain uplink routing model. It was designed to evaluate the performance in terms of energy efficiency of different uplink multi-hop routing approaches in LPWANs \cite{barrachina2016multi}. Nodes in DRESG (GW and STAs) are provided with the same selectable hardware features (i.e., transceiver, antennas etc.). Besides, every STA generates its payload and sends it to its parent (i.e., other STA or GW), which aggregates its own and all the payloads received from its direct children. Different STAs may end-up with different optimal transmission power and data rate levels depending on their distance to the GW and the number of packets to transmit. %Thus, the number of packets to be sent by an STA depends on the amount of payloads received and the number of payloads that fit in a packet $n_{p}^{\text{max}}$.
	
	\subsection{Structures and topologies}
	
\begin{figure}[h!]
		\centering
		\includegraphics[width=0.48\textwidth]{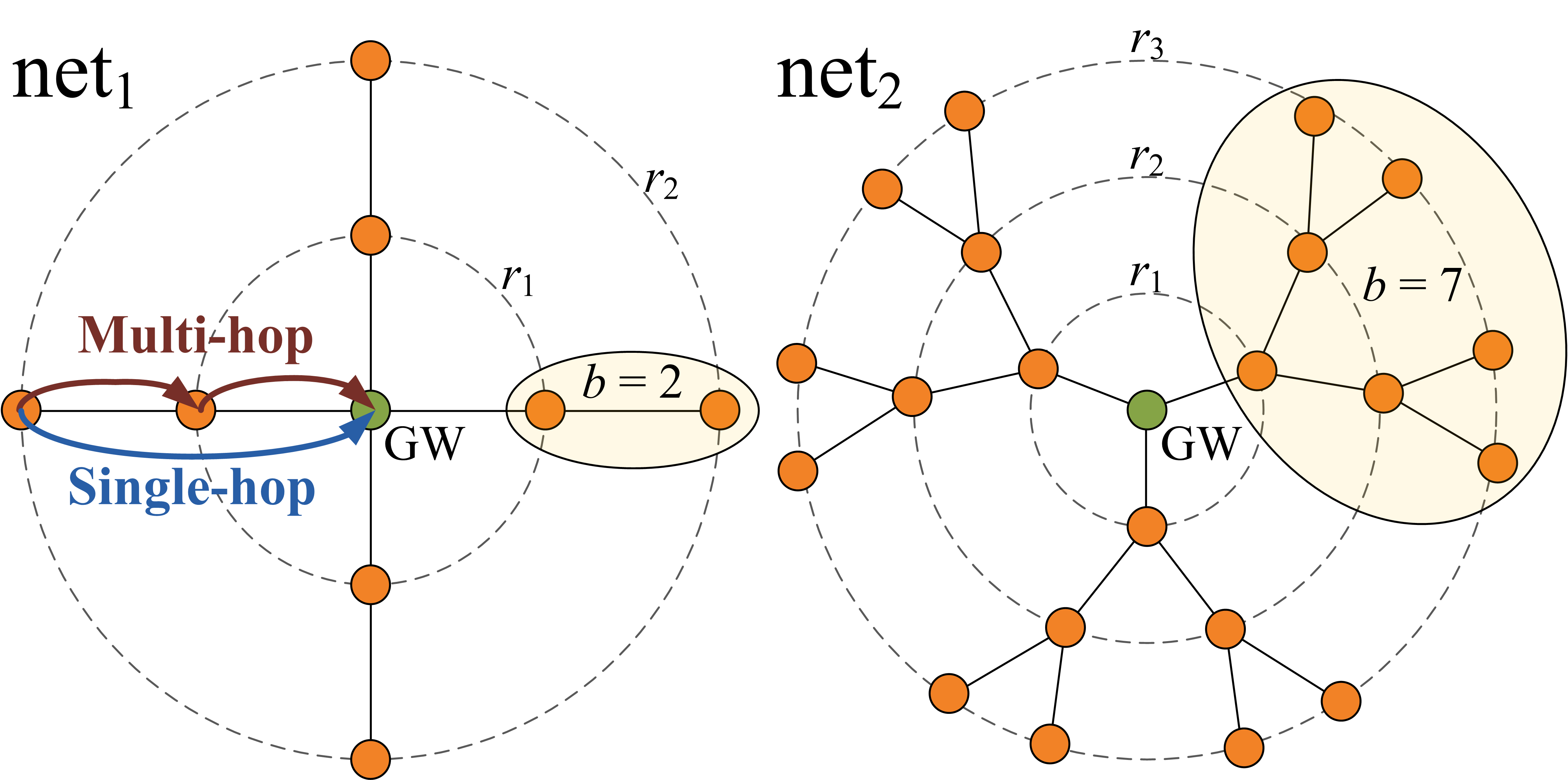}
		\caption{DRESG network structures examples. $\text{net}_1$ = \{$D$, $R = 2$, $c = 1$, $B = 4$\} and $\text{net}_2$ = \{$D$, $R = 3$, $c = 2$, $B = 3$\}. The \textit{branch loads} of networks $\text{net}_A$ and $\text{net}_B$ are 2 and 7, respectively.}
		\label{fig:net_topo}
	\end{figure}

	STAs in DRESG are spread in \textit{distance-rings} composing a tree-based network structure (see Figure \ref{fig:net_topo}) that can be defined by the following 3 parameters:
	
	\begin{itemize}
		
		\item\textbf{Number of rings ($\boldsymbol{R}$):} STAs belonging to the same ring are located at the same distance to the GW. The furthest (or last) ring is placed at distance $D$, which is given by the theoretical coverage range provided by the GW's transceiver at maximum transmission power ($P_{\text{tx}}$) and minimum data rate ($s_{\text{tx}}$). The distance among rings is defined by the spreading model, which in this work has been considered to be equidistant, i.e., same distance between any consecutive rings.
		
		\item \textbf{Tree children ratio ($\boldsymbol{c}$):} number of \textit{tree children}\footnote{DRESG distinguishes between \textit{tree children} and \textit{topology children}. On the one hand, \textit{tree children} refers to all STAs of an adjacent higher ring from which an STA (i.e., \textit{tree parent}) may receive packets. On the other hand, \textit{topology children} (children from now on) refer to the STAs in lower adjacent or non-adjacent rings from which an STA actually receives packets. Similarly, \textit{topology parent} (parent from now on) refers to that STA to which a child actually transmits its own packets (after aggregating the ones from its own children) in its way to the GW.} of every STA which does not belong to the last ring. STAs belonging to the last ring have no \textit{tree children}. The tree children ratio refers only to the network structure and it is independent of the topology (or routing). As shown in net$_1$ of Figure \ref{fig:net_topo}, different topologies (e.g., single-hop or multi-hop) may exist for the same DRESG structure.
		
		\item \textbf{Number of branches ($\boldsymbol{B}$):} a branch is a set of nodes composed of an STA in the \nth{1} ring and its direct and indirect \textit{tree children}. The node load of a branch, or \textit{branch load} ($b$), is defined as the number of STAs in a branch. Hence, the number of STAs in an DRESG network is given by $N = B\sum_{r=1}^{R}c^{r-1}$, being $c^{r-1}$ the number of nodes per branch in ring $r$ for all the branches. 
			
	\end{itemize}
	
	\subsection{Energy consumption modeling}
	
	DRESG is focused exclusively on analyzing the impact of routing in the energy consumption regardless of the medium access control (MAC) layer specification. Therefore, as a simple and ideal time division multiple access (TDMA) MAC is assumed \cite{barrachina2016multi}, the energy consumption of an STA is defined as the sum of the transmitting (TX) and receiving (RX) energies, respectively, i.e., $e = e_{\text{tx}} + e_{\text{rx}}$, being $e_{\text{tx}} =  \sum_{l=1}^{l_{\text{max}}}t_{\text{\text{tx}},p}I_{\text{tx}}(l)V_{\text{DD}}$, and $e_{\text{rx}} = t_{\text{rx}} I_{\text{rx}}V_{\text{DD}}$, where $t_{\text{rx}}$ is the time period the STA is in RX state, and $I_{\text{rx}}$ is the corresponding current consumption. The time and current consumption in TX state at power level $l$ are defined by $t_{\text{tx},l}$ and $I_{\text{tx}}(l)$, respectively. The nominal voltage is represented by $V_{\text{DD}}$.
	
	Regarding the transmission power and data rate of the children-parent connections, DRESG defines a \textit{transmission configuration} as the ordered pair $(P_{\text{tx}},s_{\text{tx}})$. Depending on the node's transceiver, one or more transmission power and data rate levels may be selectable (i.e., programmable), being the maximum data rate dependent on the receiver's sensitivity. For any given child-parent connection in the network, DRESG identifies its optimal (i.e., less energy consuming) \textit{transmission configuration} based on the distance between child and parent. 
	
	\subsection{Routing models}
	
	Due to the fact that in DRESG all the nodes in a ring have their parent in the same ring destination, three main routing models are possible:
	
	\subsubsection{Single-hop}
	all STAs (regardless of their ring $r$) transmit their data packets directly to the GW, placed at ring 0. The bottlenecks (i.e. STAs consuming most energy) are expected to be in the last ring ($r=R$) because they may require higher power levels in order to reach the GW in just one hop. As in single-hop there are no parents in the network, no RX energy is consumed.
	
	\subsubsection{Next-ring-hop}
	each node in ring $r$ transmits its data packet (or packets) to a parent node placed in the adjacent lower ring, $r-1$. The required energy transmission for a single packet is expected to be lower than in single-hop due to distances between source and destination nodes are smaller and lower transmission power levels may be used. However, payload aggregation and listening to children packets can easily generate important bottlenecks at STAs in rings close to the GW, specially in ring $r = 1$, due to the fact that these STAs aggregate all the payloads of one branch, increasing the RX and TX times in most of the cases.
	
	\subsubsection{Multi-hop (generalization)}
	each STA in ring $r$ transmits its data packet (or packets) to a parent in ring $r - \vec{\delta}(r)$, where $\vec{\delta}(r)$ is a vector of $R$ elements representing the hop length of each ring, i.e., the number of rings separating the source and destination nodes. In single-hop, $\vec{\delta}(r) = r$ for every ring, as all STAs transmit directly to the GW. Instead, in next-ring-hop, $\vec{\delta}(r) = 1$ for every ring, as all the nodes transmit to a parent in the previous ring. For instance, the single-hop and multi-hop topologies of $\text{net}_1$ in Figure \ref{fig:net_topo} can be represented by (1, 2) and (1, 1), respectively. The set of possible ring hops combinations ($\varDelta$) is given by the number of rings in the network, i.e., $|\varDelta| = R!$.
	
	\section{Learning the optimal routing} \label{sec:learning} \label{sec:learning_optimal_routing}
	
	% http://www.tokic.com/www/tokicm/publikationen/papers/KI2011.pdf
	The problem of identifying the optimal routing can be classified as an \textit{exploration/exploitation dilemma} \cite{sutton1998reinforcement} because of the implicit trade-off between the need to gain new knowledge and the need to exploit that knowledge. Specifically it is a finite-horizon multi-armed bandit problem because of the need to maximize the lifetime of STAs counting with limited capacity batteries. On the one hand, too much exploration prevents from maximizing the short-term reward because selected routing paths (i.e., actions) may entail high bottleneck consumption. But on the other hand, exploiting partial knowledge prevents from maximizing the long-term reward because already explored actions may not be the ones leading to minimum energy consumption. In this work we assume that the GW knows the full LPWAN deployment, and that it is able to estimate the energy consumed by every STA for any given routing\footnote{In most of LPWAN envisioned applications, STAs are required to be placed in the
	same location and not move over time (e.g., agriculture monitoring), facilitating the GW to gather network information such as nodes' location, routing paths, transmission configurations, etc.}.

	\subsection{The $\epsilon$-greedy approach}
	
	% http://www.tokic.com/www/tokicm/publikationen/papers/KI2011.pdf
	The $\epsilon$-greedy method \cite{watkins1989learning} tunes the randomness in action selection through a parameter $\epsilon$ that determines the probability of exploring rather than exploiting. Some of the advantages of $\epsilon$-greedy in front of other methods are its simplicity and the fact that no memorization of exploration specific data (e.g., counters or confidence bounds) is required \cite{tokic2011value}. Nonetheless, even though it is often hard to outperform \cite{vermorel2005multi}, a drawback of $\epsilon$-greedy is the fact that it is not trivial to determine which is the optimal setting (i.e., initial value and updating function) of $\epsilon$ for a given learning problem. With respect to DRESG, as network deployments are invariable along the simulations, the reward or payoff ($p$) provided by any possible routing or \textit{hops combination} ($\vec{\delta}$) defined by an action ($a$) is deterministic. Hence, exploring each action once is enough for learning its generated reward. The general $\epsilon$-greedy algorithm for learning the optimal routing in DRESG can be summarized in the steps below:
	
	\begin{enumerate}
		\item Start at iteration $i = 1$. Pick action $a_i \in \Delta$ uniformly at random. Set $\epsilon$ to its initial value $\epsilon_i=\epsilon_o$. \label{step1}
		\item Compute the reward $p(a_i) = 1/e_b(a_i)$, where $e_b(a_i)$ is the energy consumed by the most consuming (i.e., bottleneck) STA when applying action $(a_i)$. Add action $a_i$ to action history $\vec{a}$. \label{step2}
		\item New iteration: $i=i+1$. With probability $\epsilon$ pick uniformly at random a non-explored action in $\Delta$, and with probability $(1 - \epsilon)$ pick the explored action $a_i^* \in \vec{a}$ providing the maximum payoff until the current iteration. Update $\epsilon$ according to the updating function:
		\begin{itemize}
			\item Constant: $\epsilon$ remains constant throughout all the iterations, i.e., $\epsilon_i = \epsilon_o$ for $\forall i$. That is, the proportion between exploration and exploitation is kept until all actions are explored, being the best one picked since then. For instance, $\epsilon_{\text{cnt}} = 1$ would be an specification of the \textit{first explore and then exploit} approach.
			\item Decreasing: $\epsilon$ is decreased as the experiment progresses providing exploratory behavior at the beginning and exploitative behavior at the end. Selecting a proper decreasing function is not trivial due to the dependence on the feasibility of finding the optimal solution. In this work we have chosen a quadratically decreasing function, $\epsilon_i = \min \{1, \sqrt[]{\epsilon_{i-1}/i}\}$, to start exploring relatively soon due to the fact that critical lifetime constraints of STAs prevents from long exploration phases.
		\end{itemize}
		\item If the maximum number of iterations ($I$) set for the experiment has not been reached (i.e., $i<I$), go to step \#\ref{step2}. Finish otherwise.
	\end{enumerate}

	\subsection{Similarity-enhanced two layer $\epsilon$-greedy}
	
\begin{figure}[h!]
		\centering
		\includegraphics[width=0.42
		\textwidth]{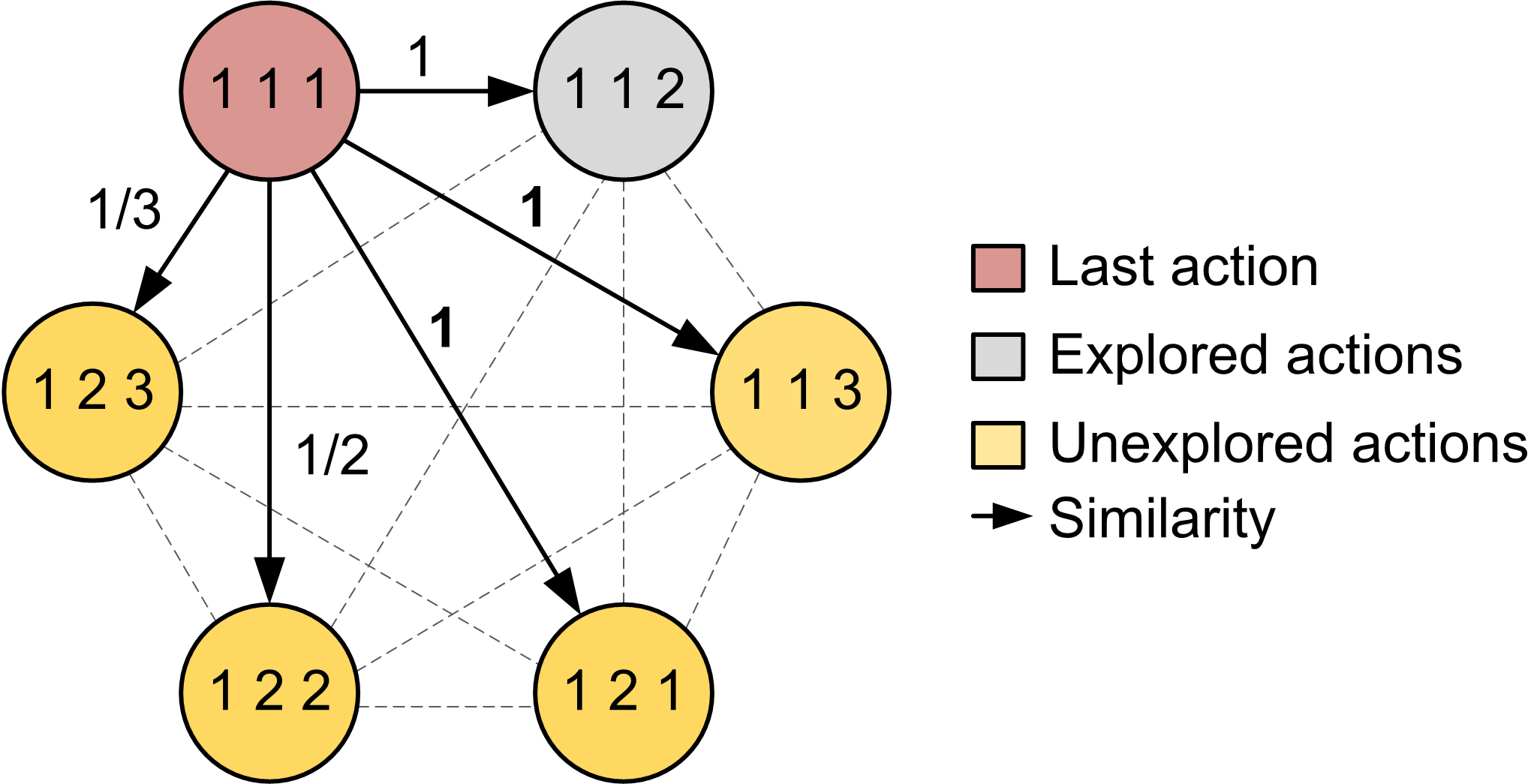}
		\caption{Example of the \textit{similarity-enhanced} two layer $\epsilon$-greedy for a DRESG network with $R=3$ rings. The algorithm should pick between action (1 1 3) and (1 2 1) with same probability because they provide maximum \textit{similarity}.}
		\label{fig:epsilon_similarity}
	\end{figure}

	As mentioned before, rewards in DRESG networks are deterministic, which allows determining with certainty which of the already explored routing paths is the most energy-efficient. In $\epsilon$-greedy, when it comes to exploring, no knowledge is used but unexplored actions are picked randomly instead. With this two layer $\epsilon$-greedy we aim to harness the gathered knowledge by exploring paths that are \textit{similar} to the best known one, expecting that they are approximately the same energy-saving. To that aim, we define the \textit{similarity} of two actions $a_i$ and $a_j$ as the inverse of the sum of the elements of the vector $d_{a_i,a_j} = |\vec{\delta(a_i)} - \vec{\delta(a_j)}|$, i.e., $S(a_i, a_j)= \big(\sum_{k} d_{a_i,a_j}(k)\big)^{-1}$ (see Figure \ref{fig:epsilon_similarity}). Specifically, in the exploring iterations, a similarity randomness setting ($\epsilon_s$) determines the probability to pick one of the unexplored actions that are most similar to the known most energy-efficient action $a_i^*$. Instead, with probability ($1 - \epsilon_s$) a random exploration is performed in order to avoid falling in local minimums.
	
	\section{Performance evaluation} \label{sec:results}
	
	% Scenarios and parameters
	In this section we study the performance of the presented learning algorithms when applied to several DRESG networks. The parameters of each scenario are obtained from the experiments presented in \cite{barrachina2016multi}. Specifically, the Texas Instruments CC1200 transceiver operating at 868 MHz, and an outdoor path loss model for 802.11ah deployments defined in \cite{hazmi2012feasibility} are assumed. Regarding data packets, the parameters implemented in the ENTOMATIC EU-project\footnote{ENTOMATIC is an agriculture plague-tracking system that intends to fight the olive fruit fly. It relies on LPWANs where STAs periodically reporting information on pest population density are spread over large olive orchards. Detailed information about the project can be found in the ENTOMATIC main website: \url{https://entomatic.upf.edu/}} are used, with data payload and header sizes of 15 and 2 bytes, respectively. The fixed packet size was set to 65 bytes, allowing to aggregate a maximum number of 4 payloads per packet.
	
	\subsection{Exploring vs. exploiting with $\epsilon$-greedy}
	
\begin{figure}[h!]
		\centering
		\includegraphics[width=0.48\textwidth]{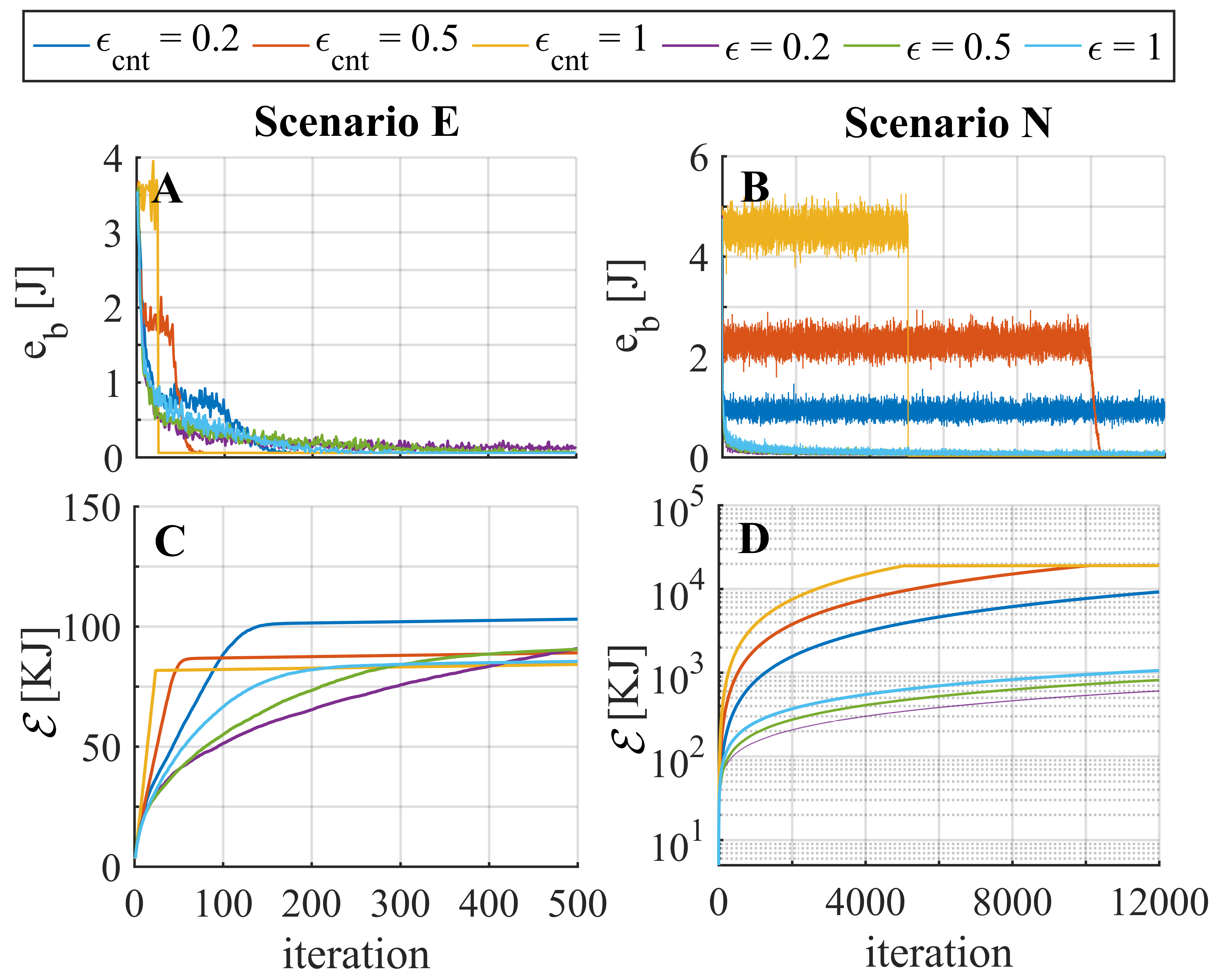}
		\caption{Empirical mean bottleneck consumption ($e_b$) and cumulated energy consumption of the \textit{historic bottleneck} ($\mathcal{E}$).}
		\label{fig:E_and_N_botle_cum}
	\end{figure}

\begin{figure}[h]
		\centering
		\includegraphics[width=0.48\textwidth]{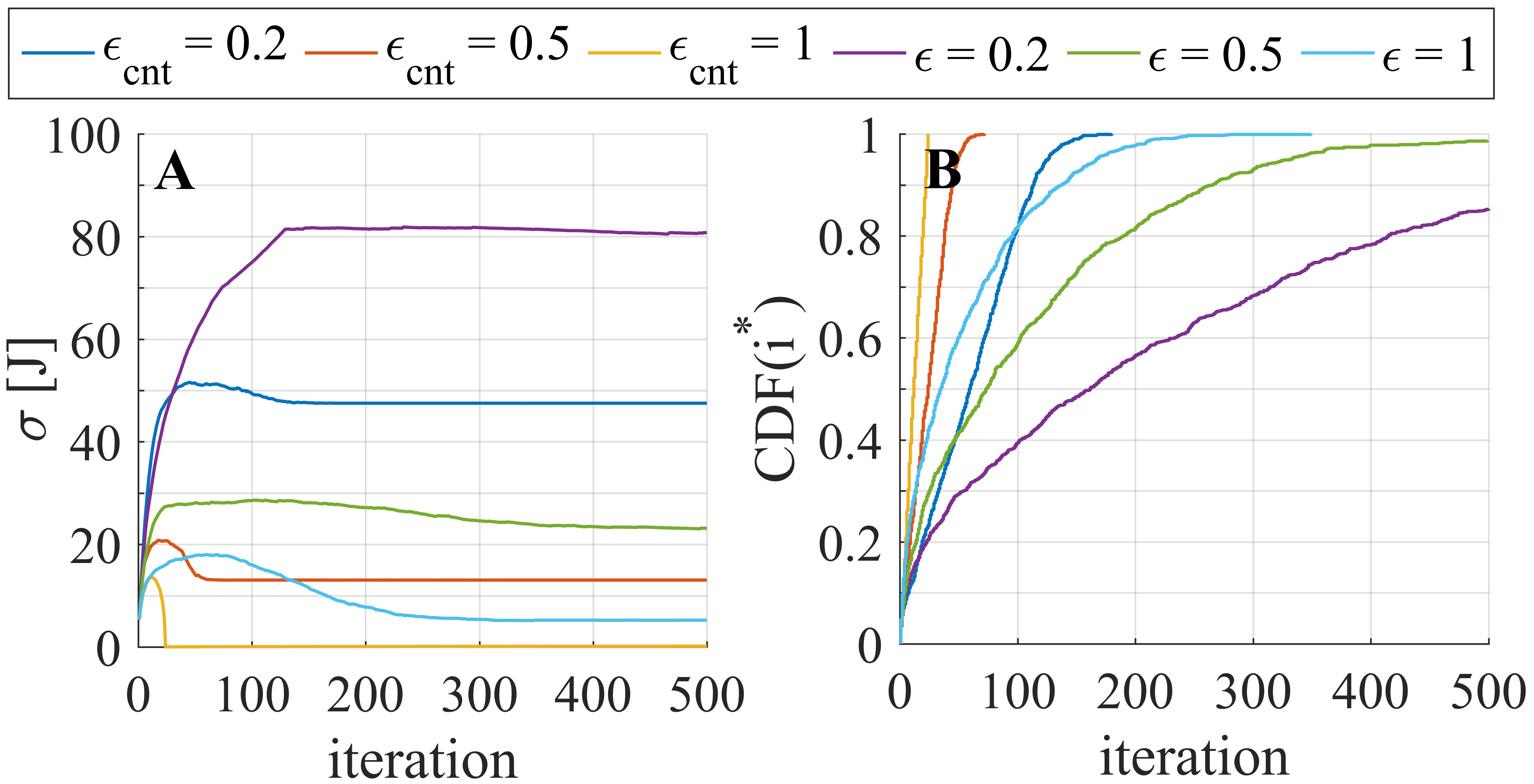}
		\caption{Empirical standard deviation of $\mathcal{E}$ ($\sigma$) and cumulative distribution function (CDF) of the iteration where the optimal action is explored ($i^{*}$) in scenario E.}
		\label{fig:std_cdf_r4_c8}
	\end{figure}

	In order to discern what is the effect of different initial values and updating functions\footnote{If not explicitly mentioned, quadratically decreasing updating functions are considered. Otherwise, for constant updating functions we use $\epsilon_{\text{cnt}}$.} of $\epsilon$-greedy, two DRESG networks, E ($R=4, c=8$) and N ($R=7, c=3$), are analyzed. In Figure \ref{fig:E_and_N_botle_cum}, the average\footnote{All the results presented in this work have been gathered through 1,000 experiment observations per scenario.} energy consumed by the bottleneck in each iteration ($e_b$), and the average energy consumed by the \textit{historic bottleneck}, i.e., the STA that has consumed more energy until a given iteration, ($\mathcal{E}$) are plotted. As expected, the curves corresponding to the \textit{first-explore-all} strategy (i.e., $\epsilon_{\text{cnt}}=1$), are the ones providing higher $e_b$ both in E and N until exploring all the actions (i.e., $|\Delta_\text{E}| = 24$ in E, and $|\Delta_\text{N}| = 5040$ in N). Since then, the optimal action is picked, reducing $e_b$ to the minimum. On the other hand, curves $\epsilon=0.2$ are the less exploring ones, as they start exploiting the best known actions frequently from the very beginning, providing in scenario E less $e_b$ on average but surpassing  $\mathcal{E}$ in the long run. Instead, in scenario N, the high number of possible actions prevents from applying \textit{first-explore-all}-like strategies due to the high consumption accumulated until starting exploiting regularly. Therefore, as the main objective of the algorithm should be to minimize $\mathcal{E}$ while taking into account the finite horizon imposed by STAs' lifetime, we note that the optimal $\epsilon$-greedy setting is really tied to the scenario (i.e., number of possible actions, differences among the generated rewards, number of STAs, etc.). Besides, even though a given randomness setting may be the best on average, as shown in Figure \ref{fig:std_cdf_r4_c8}A, the variability when few exploration is ensured (e.g., $\epsilon=0.2$) may generate infrequent but worst cases of $\mathcal{E}$, which involves an implicit risk that must be considered. In this respect, the larger the number of possible actions, the longer it takes to explore the optimal one (see Figure \ref{fig:std_cdf_r4_c8}B).
	
	\subsection{Similarity harness impact}
	
	The similarity improvement ratio is defined as the relation between the cumulated energy consumption of the \textit{historic bottleneck} at iteration $i$ when not applying similarity and when doing so for a given setting of $\epsilon$, i.e.,  $\rho_s(i) = \mathcal{E}(i)/\mathcal{E}_s(i)$. That is, whenever $\rho_s > 1$, applying similarity improves the regular $\epsilon$ setting. In the results below we consider both $\epsilon$ and $\epsilon_s$ to have the same quadratically decreasing updating function for analysis simplicity. 
	
\begin{figure}[h!]
		\centering
		\includegraphics[width=0.48\textwidth]{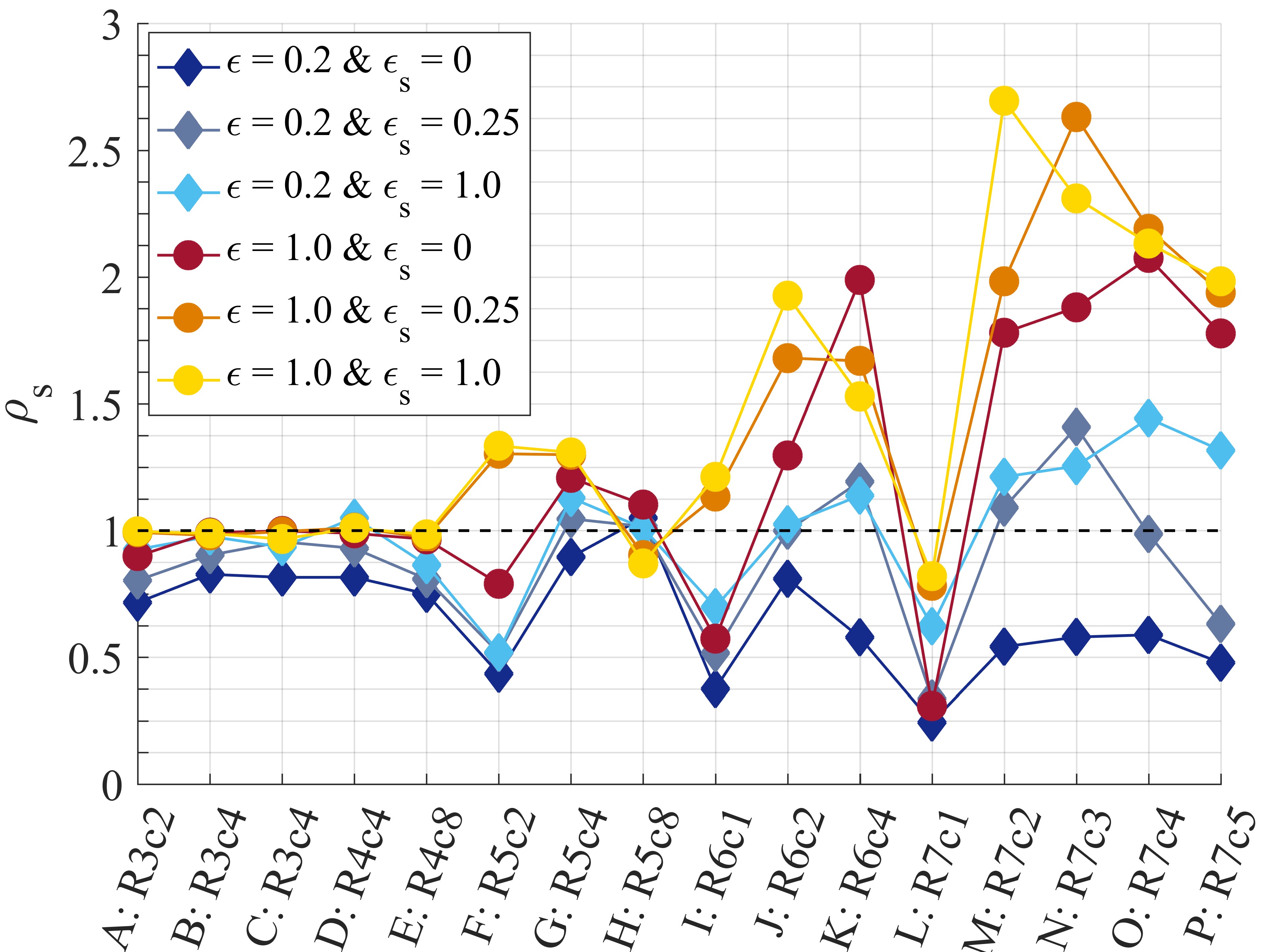}
		\caption{Similarity improvement ratio of several DRESG networks. Notation \textit{RXcY} represents networks composed of $R=X$ rings and $c=Y$ tree children ratio.}
		\label{fig:similarity_ratio_graph}
	\end{figure}

	In Figure \ref{fig:similarity_ratio_graph}, a graphical representation of the similarity consumption ratios of several DRESG networks are shown for three initial values of $\epsilon_s$. We can conclude that for networks with a small amount of possible arms (e.g., scenarios A to E with 3 and 4 rings), similarity does not improve consumption, but instead makes it worse no matter the initial values of $\epsilon$ and $\epsilon_s$. However, as networks get larger, some combinations $\{\epsilon, \epsilon_s\}$ are really energy saving. Specifically, for all the $\epsilon_s$ initial values analyzed, only the less exploring combination $\{\epsilon = 0.2, \epsilon_s = 0\}$ performs worse than with no similarity in all the scenarios. That is due to the fact that picking similar paths from the very beginning ($\epsilon_s = 0$), is giving unfavorable results as not \textit{enough} exploration is performed ($\epsilon = 0.2$). On the other hand, we note that if \textit{enough} exploration is ensured ($\epsilon = 1$), similarity can help to deeply reduce consumption (see max-exploration approach $\{\epsilon = 1, \epsilon_s = 1\}$ effect on scenario M). Also, we note that increasing the tree children ratio ($c$) generates larger $e_b$ differences among the possible routing paths, deepen the importance of discovering energy-efficient routes (even though they are not optimal) from the beginning, and not exploring much after finding them. For instance, in scenarios J to L and N to P too much exploration leads to $\rho_s$ reduction. Besides, it is remarkable the poor, and sometimes counterproductive, effect of applying \textit{similarity} in line networks (i.e., $c=1$) with large number of rings (see scenarios I and L), caused by the huge differences among the optimal routing (i.e., next-ring-hop in this case) and the rest.
	
\begin{figure}[h!]
		\centering
		\includegraphics[width=0.48\textwidth]{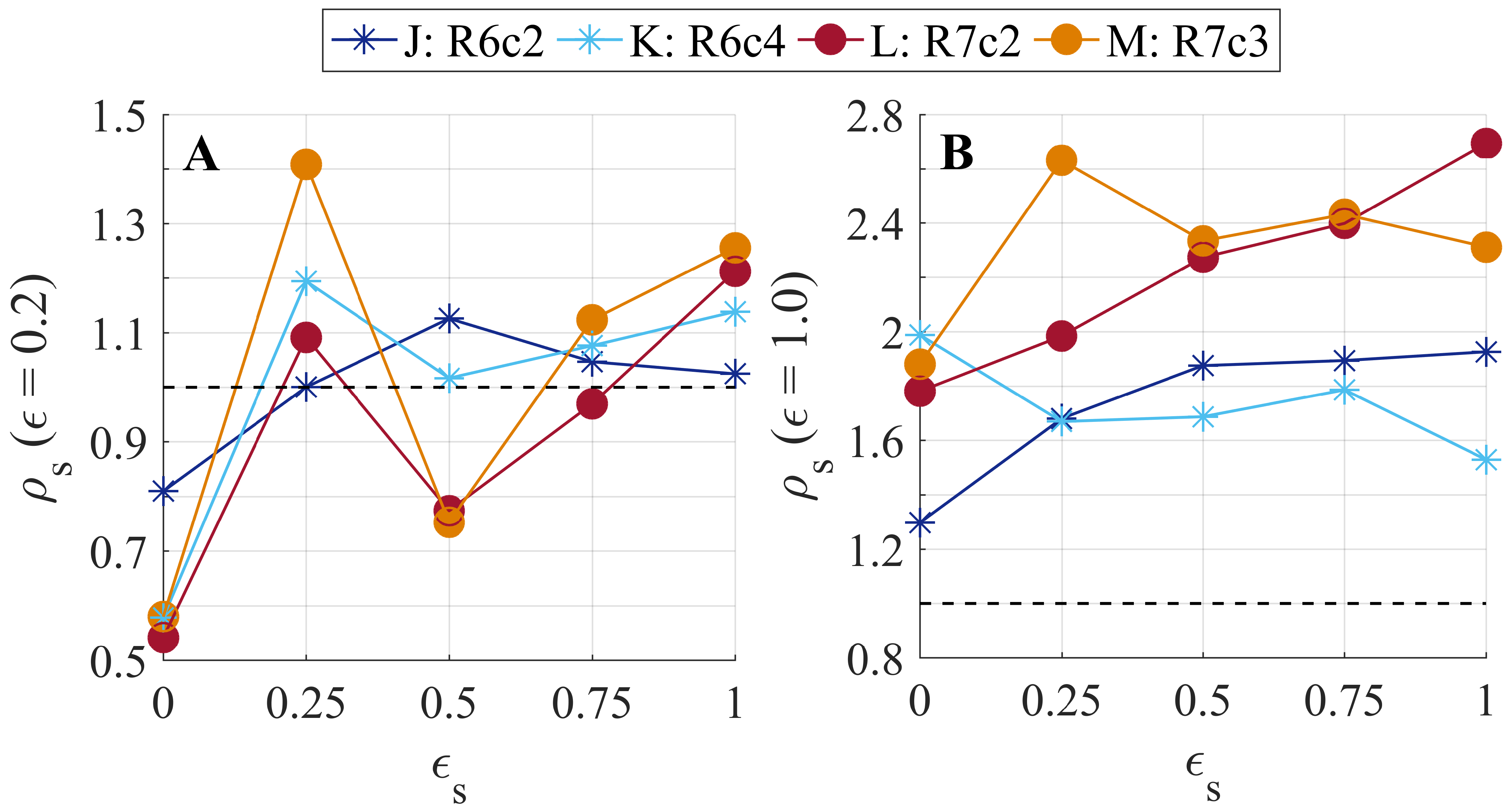}
		\caption{Effect of \textit{similarity} randomness settings on the similarity improvement ratio.}
		\label{fig:epsb_vs_similarity}
	\end{figure}

	In Figure \ref{fig:epsb_vs_similarity} we note that applying similarity when exploring very little (e.g., $\{\epsilon=0,\epsilon_s=0\}$) is not convenient if $\Delta$ is large. Also, over-exploring instead of picking similar paths may not be efficient if not \textit{enough} exploration has been performed previously (e.g., scenarios L and M for $\{\epsilon=0.2, \epsilon_s = 0.5\}$). However, in general, retarding a deep similarity exploitation (i.e., $\epsilon_s = 1$) is a good practice because of the similarity ratios provided, even though it may not be optimal. In fact, when the tree children ratio increases, if enough exploration is ensured at the beginning of the simulation (i.e, $\epsilon=1$), exploiting similarity before than with $\epsilon_s = 1$ improves the energy savings (e.g, scenarios K and M for $\epsilon=1$). Nonetheless, we see that such improvement behavior depends heavily on the DRESG network, hindering the knowledge extraction and leaving open the issue of setting the optimal initial value and updating function of both randomness tunning parameters $\epsilon$ and $\epsilon_s$.
	
	\section{Conclusions and future work} \label{sec:conclusions}
	
	Results have shown that the optimal setting of the $\epsilon$-greedy algorithm for identifying the most energy-efficient routing really depends on the LPWAN deployment. While in networks with a small number of possible routing paths it is usually better to first completely explore and then exploit the best one, for networks with larger number of rings, it is the opposite. Besides, we have shown that applying best known path \textit{similarity} can improve remarkably the energy savings in medium-large size networks if enough exploration is ensured. Tying the estimation of the optimal initial value and updating function of $\epsilon$ and $\epsilon_s$ with known DRESG network parameters, such as the number of rings and the tree children ratio, is an open issue to deal with in future works.
	
	\section*{Acknowledgment}	
	This work has been partially supported by the Spanish Ministry of Economy and Competitiveness under the Maria de Maeztu Units of Excellence Programme (MDM-2015-0502), and by the ENTOMATIC FP7-SME-2013 EC project (605073).
	
	\bibliographystyle{unsrt}
	\bibliography{bib}

\end{document}